\documentclass{aa}
\usepackage{amstext}
\usepackage{graphicx}
\usepackage{hyperref}

\begin{document}

\titlerunning{Relativistic electrons} 
\authorrunning{Mann et al.}

\title{Generation of relativistic electrons at the termination shock in the solar flare region} 

\author{G.~ Mann\inst{1}, A. M. Veronig\inst{2}, \and F.~Schuller\inst{1}} 
\institute{Leibniz-Institut f\"ur Astrophysik Potsdam, An der Sternwarte 16, 
    D-14482 Potsdam, Germany; GMann@aip.de
\and
    Institute of Physics, University of Graz, Universit\"atsplatz 5, 
        A-8010 Graz, Austria}

\offprints{G.~Mann, \email{GMann@aip.de}}

\date{Accepted: April 2024}

\abstract
{Solar flares are accompanied by an enhanced emission of electromagnetic waves
from the radio up to the $\gamma$-ray range. The associated hard X-ray and microwave 
radiation is generated by energetic electrons. These electrons play an important role, 
since they carry a substantial part of the energy released during a flare. The flare is 
generally understood as a manifestation of magnetic reconnection in the corona.
The so-called standard CSHKP model is one of the most widely accepted models for eruptive flares.
The solar flare event on September 10, 2017 offers us a unique opportunity to study
this model. The observations from the Expanded Owens Valley Solar Array (EOVSA)
show that $\approx$ 1.6 $\times$ 10$^{4}$ 
electrons with energies $>$ 300~keV are generated in the flare region.}
{There are signatures in solar radio and extreme ultraviolet (EUV) observations
as well as numerical simulations 
that a “termination shock” (TS) appears in the magnetic reconnection outflow region. 
Electrons accelerated at the TS can be considered to generate the loop-top hard X-ray sources.
In contrast to previous studies, we investigate whether the heating of the plasma at 
the TS provides enough relativistic electrons needed for the hard X-ray and microwave emission 
observed during the solar X8.2 flare on September 10, 2017.} 
{We studied the heating of the plasma at the TS by evaluating the jump in the temperature 
across the shock by means of the Rankine-Hugoniot relationships under coronal circumstances 
measured during the event on September 10, 2017. The part of relativistic electrons
was calculated in the heated downstream region.}
{In the magnetic reconnection outflow region, the plasma is strongly heated at the TS. 
Thus, there are enough energetic electrons in the tail of the electron distribution function 
(EDF) needed for the microwave and hard X-ray emission observed during the event 
on September 10, 2017.}   
{The generation of relativistic electrons at the TS is a possible mechanism of 
explaining the enhanced microwave and hard X-ray radiation emitted during flares.}


\maketitle

%
\section{Introduction}
On the Sun, a flare occurs as a sudden enhancement of the local emission of
electromagnetic waves covering the whole spectrum from the radio up to the $\gamma$-ray range
with a duration of minutes to hours (see Aschwanden (2005) as a textbook).
During flares, a large amount of stored magnetic energy is suddenly released and transferred
into the local heating of the coronal plasma, mass motions (as e.g., jets and coronal mass ejections (CMEs)), and the 
generation of energetic particles (usually called solar energetic particle (SEP) events)
(see e.g., Heyvaerts 1981; Lin 1974; Reames, Barbier, \& Ng 1996; Klein \& Trottet 2001). 
The flare is considered to be a manifestation of magnetic reconnection in the corona. 
Presently, it is widely understood in terms of the so-called standard or CSHKP model 
(Carmichael 1964; Sturrock 1966; Hirayama 1974; Kopp \& Pneumann 1976) that
was developed to explain eruptive flares; that is, flares associated with a CME. 
According to the CSHKP model, a filament is rising up due to its photospheric footpoint motions,
leading to the stretching of the underlying magnetic field lines and, subsequently, to the 
establishment of a current sheet. If the current within this sheet exceeds a critical value, 
plasma waves are excited due to different plasma instabilities, leading to an enhancement 
of the resistivity (see. e.g., Treumann \& Baumjohann (1997)). Then, magnetic reconnection 
can take place in the region of enhanced resistivity, which is called the “diffusion region.” 
Due to the strong curvature of the magnetic field lines in the vicinity of the diffusion region,
the plasma slowly inflowing into the reconnection site is shooting away from the 
diffusion region, forming the outflow region. 

The generation of energetic electrons during flares is of special interest, 
since they carry a substantial part of the energy released during flares 
(Lin \& Hudson 1971, 1976; Emslie et al. 2004, Krucker et al. 2010, Krucker \& Battaglia 2014). 
Observations with the Reuven Ramaty High Energy Solar Spectroscopic Imager (RHESSI)
(Lin et al. 2002) reveal that 10$^{36}$ electrons
with energies beyond 30~keV are typically produced per second during X-class flares 
(Warmuth et al 2007).
There are several models for electron acceleration in the present debate
(see e.g., Chapter 11 in the textbook by Aschwanden (2005) as well as
Zharkova et al. (2011) and Mann (2015) as reviews). One of them is electron acceleration
at the so-called “termination shock” (TS). If the velocity of the plasma in the outflow region 
becomes super-Alfv\'enic, a shock wave can be established.
For instance, it can happen due to the interaction of the outflow jet with the underlying loops.
Numerical simulations (Forbes 1986, 1988; Forbes \& Malherbe 1986, Workman et al. 2011;
Takasao et al. 2015; Takasao \& Shibata 2016; Takahashi et al. 2017; Shen et al. 2018; 
Kong et al. 2019; Cai et al. 2021)
support the establishment of a TS in the magnetic reconnection outflow region. 
Masuda et al. (1994), Shibata et al. (1995), and Tsuneta \& Naito (1998) reported on  
so-called loop-top sources of the hard X-ray radiation. 
Tsuneta \& Naito (1998) proposed that the TS could be the source of energetic electrons,
which are needed for the hard X-ray emission that can explain these loop-top soures. 
Solar radio (Aurass et al. 2002; Aurass \& Mann 2004; Chen etal. 2015) 
and extreme ultraviolet (EUV) (Polito et al. 2018; Cai et al. 2021) observations 
revealed signatures of such TSs in flare regions.
Mann et al. (2006, 2009) and Warmuth et al. (2009) 
discussed the generation of energetic electrons at the TS by “shock drift acceleration” (SDA).
In contrast to these studies, here we investigate whether the heating of the plasma at the TS
provides enough of the relativistic electrons needed for the microwave and 
hard X-ray emission during flares. 

To do that, we investigate the X8.2 flare that occurred on September 10, 2017.
The event took place on the western solar limb and was associated with a fast CME.
It exhibits various characteristics of a textbook event following the CSHKP scenario, 
like the formation of a hot flux rope, an indication of magnetic reconnection jets, 
and signatures of a large-scale hot current sheet beneath
the eruption  (e.g., Cheng et al. 2018, Veronig et al. 2018, Warren et al. 2018).
Therefore, this event offers us an excellent opportunity to study the flare scenario.
Chen et al. (2020) measured the plasma parameters as the electron number density,
the magnetic field, the temperature, and the number density of relativistic electrons 
in the flare region by employing microwave imaging observations with the newly
commissioned Expanded Owens Valley Solar Array (EOVSA; Gary et al. 2018). 
They found that the microwave-emitting energetic electrons are strongly concentrated
in the loop-top region. The presence of a TS there could be a possible explanation. 
Hence, these measurements allow us to determine both the plasma parameters at the TS and 
the number density of electrons with energies $>$~300~keV. These values were used to demonstrate 
that the heating of the plasma at the TS provides enough energetic
electrons needed for the microwave and hard X-ray emission during flares.


In Section~2, the properties of the TS are derived by evaluating the Rankine-Hugoniot 
relationships under special circumstances
in the flare region. The observational results measured by the EOVSA instrument 
(Gary et al. 2018) during the solar flare on September 10, 2017 (Chen et al. 2020) are presented 
in Section~3. Whether the heating of the plasma at the TS can provide enough 
of the relativistic electrons with energies
$>$ 300~keV needed for the observed microwave and X-ray radiation are discussed in Section~4.
The results of the paper are summarized in Section~5.
\section{Properties of the termination shock}
The TS is considered to be a nearly perpendicular shock (Tsuneta \& Naito 1998, Mann et al. 2006, 2009).
It is confirmed by numerical simulations (see e.g., Fig.~1e in Kong et al. (2019)). 
In magnetohydrodynamics (MHD), shock waves are described in terms of the so-called
Rankine-Hugoniot relationships (see e.g., Priest (1982); 
here, we followed the approach given in Appendix A in the paper by Mann et al. (2018)). 
The shock normal, $\vec n_{s}$, is directed along the x axis; in other words, 
the shock itself is located in the y-z plan (see Fig.~1). 
\begin{figure}[t]
   \centerline{\includegraphics[width=0.42\textwidth]{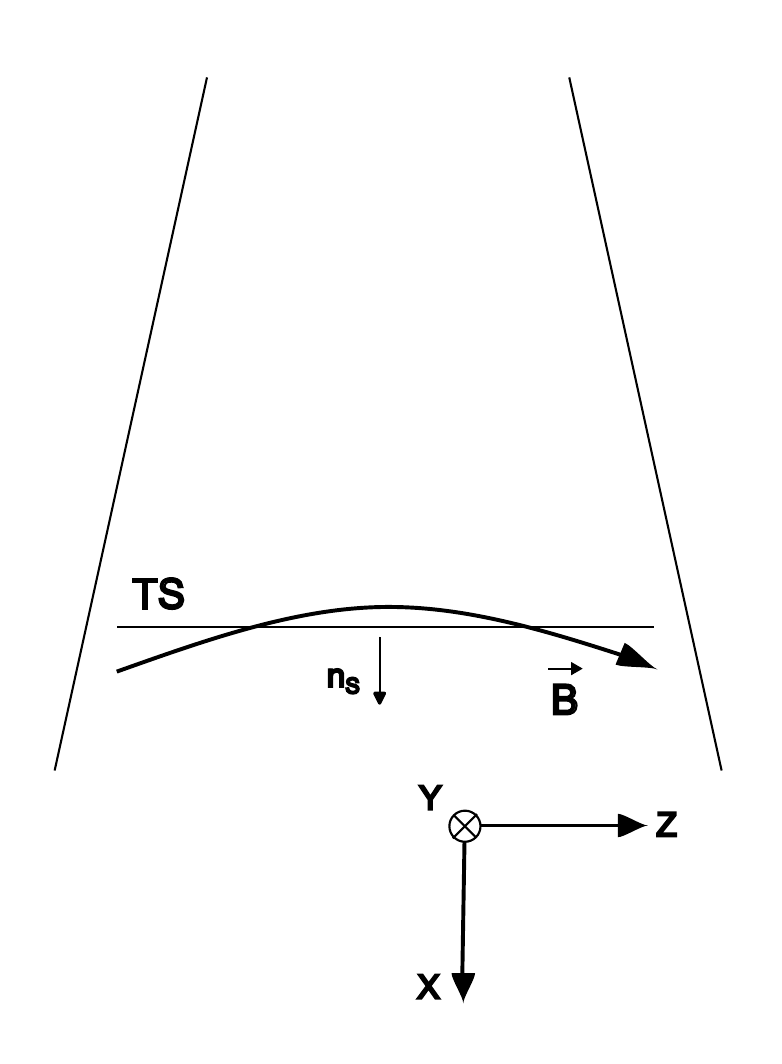}}
   \caption{Sketch of the magnetic reconnection outflow region with the TS.
           The upstream and downstream regions of the TS are located above and below the TS,
           respectively. The coordinate system employed in the paper is also drawn.}    
\label{Fig.1}
 \end{figure}
The magnetic field, $\vec B$,
is put in the x-z plane and takes an angle, $\theta$, to the x axis in the upstream region. 
All of the quantities with the subscript 1 or 2 denote those in the upstream and 
downstream regions, respectively.
These Rankine-Hugoniot relationships were evaluated for the special case of 
a nearly perpendicular, fast-mode shock -- $\theta \rightarrow$~90$^{\circ}$ -- leading to 
\begin{equation}
\frac{B_{2}}{B_{1}} = X
\end{equation}
(see Eq.~(A.9) in Mann et al. (2018) for the jump in the magnetic field) and 
\begin{equation}
\frac{T_{2}}{T_{1}} = 1 + \frac{(\gamma-1)}{2} \cdot \frac{v_{A1}^{2}}{c_{s1}^{2}} \cdot (M_{A})^{2} \cdot
\left( 1 - \frac{1}{X^{2}} \right)
\end{equation}
(see Eq.~(A.8) in Mann et al. (2018) for the jump in the temperature). 
Here, $X = \rho_{2}/\rho_{1}$ denotes the jump in the mass density across the shock. The mass density, $\rho$,
is related to the full particle number density, $N$, 
by $\rho = \tilde \mu m_{p} N$ ($m_{p}$, proton mass; $\tilde \mu$ = 0.6, mean molecular weight (Priest 1982)). 
The equation of state, $p = Nk_{B}T$ ($k_{B}$, Boltzmann's constant), relates the pressure, $p$, to the 
full particle number density, $N$, and the temperature, $T$. 
Here, the Alfv\'en velocity and sound speed are given by $v_{A1} = B_{1}/(4\pi \rho_{1})^{1/2}$ 
and $c_{s1} = (\gamma k_{B}T_{1}/\tilde \mu m_{p})^{1/2}$  
($\gamma$, ratio of the specific heats) in the upstream region, respectively. 
In the flare region, $v_{A,1} \gg c_{s,1}$ can be assumed, as is shown in Section~4. 
Then, according to the Rankine-Hugoniot relationships, 
the Alfv\'en-Mach number $M_{A} = v_{s}/v_{A1}$ of the shock is related 
to the density jump, $X$, by
\begin{equation}
M_{A} = \sqrt{X \cdot 
   \frac{ [\gamma+X(2-\gamma)]}{[(\gamma+1)-X(\gamma-1)]}}
\end{equation}
(see Eq.~(A.10) in Mann et al. (2018)) with the shock speed, $v_{s}$. 

The TS is a fast magnetosonic shock that is connected with positive jumps in the mass density,
the magnetic field, and the temperature. Now, we look to the jump of the temperature at the TS 
(see Eq.~(2)).
Crucially, it depends on the ratio $v_{A,1}/c_{s1}$. This ratio is $\gg$ 1 in flare regions, 
as is discussed in Section~4. Hence, the plasma is strongly heated from the temperature 
in the upstream region, $T_{1}$, 
up to a temperature in the downstream region of the TS, $T_{2}$, under special circumstances 
in flare regions. 
Thus, it could be that there are relativistic electrons in the tail of the electron 
distribution function (EDF) in the downstream region. 
Next, we calculated the part of electrons with energies beyond 
an energy, $E^{\ast}$, in the downstream EDF. 

To do that, we assumed a Maxwellian EDF in the downstream regions with temperature $T_{2}$:
\begin{equation}
   f_{M}(\vec p) = C_{M} \cdot e^{-E_{kin}/k_{B}T_{2}}
   .\end{equation}

Microwave and hard X-ray observations during the flare on September 10, 2017
reveal that the EDF in the loop-top region consists of a thermal core and a broken power law 
in the energetic tail (see Fig.~3c in Chen et al. (2021) and the third paragraph of Sect.~3).
Such an EDF is described by several parameters as the temperature of the thermal core as well as 
the break energy and the power law indices of the broken power law. In contrast to such an EDF, 
a Maxwellian one has only one parameter, namely the temperature.
Unfortunately, the parameters of a broken power law cannot be determined 
by macroscopic relationships such as, for example, the equation of momentum. 
Since the Rankine-Hugoniot relationships result from the macroscopic MHD equations,
these relations do not allow one to derive any information about the change 
in the parameters of the broken power law across the shock, such as the power law indices, 
for instance. For these reasons and 
for simplicity's sake, we chose a Maxwellian EDF in the downstream region, since it has only 
the temperature, $T$, as the parameter and the Rankine-Hugoniot relationships 
provide information about the jump in the temperature across the shock.

%
%

In order to derive the differential electron number density, 
we followed the approach presented in the paper by Mann (2018).
The kinetic energy, $E_{kin}$, of an electron with the momentum, $p$ 
($p = \| \vec p \|$), is given by $E_{kin} = (p^{2}c^{2}+m_{e}^{2}c^{4})^{1/2}-m_{e}c^{2}$
($m_{e}$, electron mass). The constant, $C_{M}$, can be fixed to 
\begin{equation}
\frac{1}{C_{M}} = 4\pi(m_{e}c)^{3} \cdot \int_{0}^{\infty} d\epsilon
\cdot \sqrt{\epsilon(2+\epsilon)} \cdot (1+\epsilon) \cdot 
      e^{-\epsilon/\epsilon_{th}}
\end{equation}
(see Eq.~(3.14) in Mann (2018)), with $\epsilon = E_{kin}/m_{e}c^{2}$ and 
$\epsilon_{th} = k_{B}T_{2}/m_{e}c^{2}$,
if the EDF is normalized to unity. Then, the number density, $N_{e}(\epsilon^{\ast})$,
of electrons with energies beyond $E^{\ast}$ is found to be
\begin{equation}
   \frac{N_{e}(\epsilon^{\ast})}{N_{e,0}} = 4\pi(m_{e}c)^{3} C_{M} 
      \cdot \int_{\epsilon^{\ast}}^{\infty} d\epsilon
      \cdot \sqrt{\epsilon(2+\epsilon)} \cdot (1+\epsilon) \cdot 
      e^{-\epsilon/\epsilon_{th}}
,\end{equation}
with $\epsilon^{\ast} = E^{\ast}/m_{e}c^{2}$ and $N_{e,0}$ as the complete electron
number density. Then, the differential electron number density defined by
\begin{equation}
\frac{dN_{e}(E)}{dE} = \frac{1}{m_{e}c^{2}} \cdot \frac{dN_{e}}{d\epsilon}
\end{equation}
results from Eq.~(6) as
\begin{equation}
   \frac{dN_{e}(E)}{dE} = \frac{4\pi (m_{e}c)^{3}C_{M}}{m_{e}c^{2}} 
   \cdot \sqrt{\epsilon(2+\epsilon)} \cdot (1+\epsilon) \cdot e^{-\epsilon/\epsilon_{th}}
\end{equation}
in the case of a Maxwellian EDF with temperature $T_{2}$. 
\section{Solar flare on September 10, 2017}
The solar event on September 10, 2017 was an X8.2 flare and the second-largest one in 
solar cycle~24. Since it appeared at the limb, it is an unique event  
through which to study the flare scenario in an excellent manner. 
It was described in detail from different points of view (Veronig et al. 2018; 
Warren et al. 2018; Gary et al. 2018; Longcope et al. 2018; Hayes et al. 2019; 
Morosan et al. 2019; Chen et al. 2020; Yu et al. 2020; French et al. 2020; 
Reeves et al. 2020; Fleishman et al. 2020, 2022).
This event confirms well the standard (CSHKP) flare model. 
It showed an erupting flux rope with 
a CME and an underlying current sheet. The event was also acompanied 
by an SEP event (Guo et al. 2018) and a long-lasting ($\approx$ 12 hours) $\gamma$-ray emission 
with energies $>$~100~MeV (Omodei et al. 2018).

In this event, signatures of a long hot current sheet were observed behind the erupting flux rope 
(Seaton \& Darnel 2018; Warren et al. 2018, Yan et al. 2018) in the flare region. 
This led to the formation of a large-scale reconnection current sheet (RCS). The mircrowave 
spectral imaging observations with the EOVSA instrument (Gary et al. 2018) reveals 
that relativistic electrons 
are present in the region between the erupting flux rope and the underlying flare loop arcade
(Chen et al. 2020). Inspection of Figure~2a in Chen et al. (2020)
shows that the RCS extends in the height range of 20-70~Mm in projection above the limb. 
Chen et al. (2020) identified the RCS as a bright feature in the 193~\AA~  band. 
It was observed by the “Atmospheric Imaging Assembly” (AIA, Lemen et al. 2012) instrument
on board the “Solar Dynamics Observatory” (SDO, Pesnel et al. 2012). 
The 193~\AA~ band is sensitive for hot plasmas 
with a temperature of about 18~MK. Figure~3a in Chen et al. (2020) reveals 
that the X and Y points
are located at heights of $\approx$ 32~Mm and $\approx$ 20~Mm above the limb in projection, 
respectively. 
According to numerical simulations (Forbes 1986, 1988; Forbes \& Malherbe 1986, 
Workman et al. 2011;
Takasao et al. 2015; Takasao \& Shibata 2016; Takahashi et al. 2017; Shen et al. 2018,
Kong et al. 2019), the TS is predicted in the region between the Y point and 
the underlying flare loop arcade. 
Hence, the TS is expected in the height region of about 20~Mm in the event on September 10, 2017. 
There, an enhanced hard X-ray radiation is observed. 
The hard X-ray source observed by the RHESSI spacecraft (Lin et al. 2002)
is located at the top of the flare loop arcade with a FWHM size 
of $\approx$~20''~$\times$~30'' (or 14.5~Mm~$\times$~21.8~Mm) (see Figs. (1a) and (1d)
in Chen et al. (2021)). There, the number density of electrons 
with energies beyond 300~keV is much greater, as is shown in Figure~3e in Chen et al. (2020). 
That is an indication that the TS could be the source of these relativistic electrons, 
as was proposed by Tsuneta \& Naito (1998).

The microwave data of the EOVSA instrument (Gary et al. 2018) 
and the hard X-ray data of the RHESSI instruments allow one to derive the EDF at the loop-top
source of this event. Chen et al. (2021) did that for a broad energy range, 10~keV $-$ 1~MeV,
during the early impulsive phase of the flare by means of the forward fitting method 
(Holman et al. 2003). The resulting EDF consists of a thermal core with a temperature of
$\approx$~25~MK, a broken power law in the energetic tail beyond $\approx$~16~keV,
and an electron power law index of $\approx$~3.6. Beyond $\approx$~160~keV, the power law breaks 
to a power law with an index of $\approx$~8.5 (see Fig.~3c in Chen et al. (2021)).

The upstream and downstream regions of the TS are located in the regions between the Y point 
and the TS and between the TS and the flare loop arcade, respectively. 
There, we find a temperature of $T_{1}$ = 18~MK in the upstream region (see Fig.~2a 
in Chen et al. (2020) and also Warren et al. (2018)) as well as an electron number density,
$N_{e,2}$ = 10$^{10}$~cm$^{-3}$, and magnetic field, $B_{2}$ = 400~G 
(see Fig.~3b in Chen et al. (2020)), in the downstream region as the plasma parameters at the TS.
Furthermore, the number density of electrons with energies $>$~300~keV 
is found to be $\approx$~1.6 $\times$ 10$^{4}$~cm$^{-3}$ (because 
of log[$N_{e,>300keV}(cm^{-3})$] = 4.2 in Fig.~3e in Chen et al. (2020)) below the Y point. 
\section{Discussion}
The aim of this section is to show that the heating of the plasma at the TS
provides enough relativistic electrons in the downstream region of the TS
by employing data from the solar event on September 10, 2017
(see Sect.~3).

As was mentioned in Section~3, the temperature $T_{1}$ = 18~MK (corresponding to 
an energy of 1.55~keV) is found in the upstream region of the TS. It results in
a sound speed of $c_{s,1} = (\gamma  k_{B}T_{1}/\tilde \mu m_{p})^{1/2}$ = 643~km s$^{-1}$
for $\gamma$ = 5/3. Furthermore, a magnetic field strength of $B_{2}$ = 400~G and 
an electron number density of $N_{e,2}$ = 10$^{10}$~cm$^{-3}$ are adopted in the downstream
region, leading to an Alfv\'en velocity of
$v_{A,2} = B_{2}/(4\pi \tilde \mu m_{p} N_{2})^{1/2}$ = 8128~km s$^{-1}$
for $\tilde \mu$ = 0.6 (Priest 1982).
\footnote{In the coronal plasma, the full particle number $N$ density is related
to the electron number density $N_{e}$ by $N$ = 1.92$N_{e}$ (Mann et al. 1999).}
Thus, the ratio $v_{A,2}/c_{s,1}$ $=$ 12.6 -- the assumption that
$(v_{A,2}/c_{s,1})^{2} \gg 1$ to derive Eq.~(3) -- is well justified.

Next, the ratio $T_{2}/T_{1}$ had to be calculated by means of Eq.~(2).
Inserting Eq.~(3) for $M_{A}$ in Eq.~(2) and taking into account
$v_{A,1} = v_{A,2}/X^{1/2}$, one gets 
\begin{equation}
\frac{T_{2}}{T_{1}} = 1 + \frac{1}{6} \cdot \frac{v_{A,2}^2}{c_{s,1}^{2}} \cdot \frac{(X^{2}-1)^(5+X)}{X^{2}(4-X)}
.\end{equation}
The jump in the temperature, $T_{2}/T_{1}$, across the TS increases from one to infinity
if $X$ changes from one to four. It is calculated for several values of the density jump, $X$, 
varying in the range of 1.1-1.6 by means of Eq.~(9) and $v_{A,2}/c_{s,1}$ = 12.6. 
The results are summarized
in Table~1. Additionally, the electron number density, $N_{e,1}$, the magnetic field, $B_{1}$,
and the Alfv\'en velocity, $v_{A,1}$, of the upstream region are also listed in Table~1. 
\begin{table*}[t]
   \caption{Parameters of the TS according to the Rankine-Hugoniot relationships for 
            density jumps, $X$, in the range of 1.1-1.6 for the plasma conditions 
            measured during the solar event on September 10, 2017,
            i.e., the temperature, $T_{1}$ = 18~MK, in the upstream region as well as 
            the electron number density, $N_{e,2}$ = 10$^{10}$~cm$^{-3}$, and 
            the magnetic field, $B_{2}$ = 400~G, in the downstream region. 
            The Alfv\'en-Mach number, $M_{A}$ (second column), was calculated with Eq.~(3).
            $N_{e,1} = N_{e,2}/X$ (third column), $B_{1} = B_{2}/X$ (fourth column),
            and $v_{A,1} = v_{A,2}/X^{1/2}$ (fifth column) are obtained for the electron number density,
            the magnetic field, and the Alfv\'en velocity, respectively, in upstream regions of the TS. 
            The temperature in the downstream region, $T_{2}$, was determined by 
            Eq.~(9). The corresponding thermal energies, $k_{B}T_{2}$, are given in the sixth column.
            In the upstream region, a temperature of $T_{1}$ = 18~MK leads to a thermal energy of
            $k_{B}T_{1}$ = 1.55~keV. 
            $N_{e,2>300keV}/N_{e,2}$ (seventh column) and $N_{e,2>20keV}/N_{e,2}$ (eighth column)
            denote the ratios between the number density of electrons with energies 
            beyond 300~keV and 20~keV, respectively, and the full electron number density 
            in the downstream region, $N_{e,2}$. They were calculated by means of Eq.~(6),
            with $\epsilon_{300keV}^{\ast}$ = 0.5859 and $\epsilon_{20keV}^{\ast}$ = 0.0391.}
      \begin{center}
   \begin{tabular}[h]{cccccrcc} \hline 
   $X$ & $M_{A}$ & $N_{e,1}(10^{10}cm^{-3})$ & $B_{1}$(G) & $v_{A,1}$(km s$^{-1})$ & $k_{B}T_{2}$(keV) & $N_{e,2>300keV}/N_{2}$ & $N_{e,2>20keV}/N_{e,2}$ \\
   \hline 
   1.10 &  1.076 & 0.9091 & 364 & 7757 &  16.7 & 5.451 $\times$ 10$^{-8}$ & 0.511 \\
   \\            
   1.11 &  1.083 & 0.9009 & 360 & 7707 &  18.0 & 4.870 $\times$ 10$^{-7}$ & 0.554 \\
   1.12 &  1.091 & 0.8929 & 357 & 7677 &  19.4 & 1.536 $\times$ 10$^{-6}$ & 0.578 \\
   1.13 &  1.099 & 0.8850 & 354 & 7646 &  20.7 & 3.871 $\times$ 10$^{-6}$ & 0.606 \\
   \\
   1.20 &  1.153 & 0.8333 & 333 & 7412 &  29.5 & 2.497 $\times$ 10$^{-4}$ & 0.737 \\
   1.30 &  1.232 & 0.7692 & 308 & 7136 &  41.1 & 3.677 $\times$ 10$^{-3}$ & 0.830 \\
   1.40 &  1.313 & 0.7143 & 286 & 6876 &  51.6 & 1.444 $\times$ 10$^{-2}$ & 0.876 \\
   1.50 &  1.396 & 0.6667 & 267 & 6645 &  61.5 & 3.360 $\times$ 10$^{-2}$ & 0.904 \\
   1.60 &  1.483 & 0.6250 & 250 & 6426 &  70.9 & 5.971 $\times$ 10$^{-2}$ & 0.923 \\
   \hline
   \end{tabular}
   \end{center}
   \end{table*}
The sixth column reveals that the plasma is strongly heated across the TS.
The thermal energies in the downstream region, $k_{B}T_{2}$, are given in the sixth column 
of Table~1. They vary from 16.7~keV to 70.9~keV for a TS with Alfv\'en-Mach 
numbers, $M_{A}$, in the range of 1.10-1.60, respectively. 

Since the plasma is heated in the downstream region, a Maxwellian EDF was assumed there
for simplicity's sake (see also the discussion in Sect.~2).
The number density of electrons with energies beyond 300~keV, $N_{e,2>300keV}$, 
was calculated with Eq.~(6) for different values of the density jump, $X$, in the range of 1.1-1.6, as is presented
in the seventh column in Table~1.
We note that the EDF was treated in a fully relativistic manner in Section~2.  
The measurements with EOVSA (Gary et al. 2018) during the solar event on September 10, 2017 
provide $N_{e,2>300keV}$ = 1.58 $\times$ 10$^{4}$~cm$^{-3}$ (see Section 3 and Fig.~3e in Chen et al. (2020))
or $N_{e,2>300keV}/N_{e,2}$ = 1.58 $\times$ 10$^{-6}$ for $N_{e,2}$ = 10$^{10}$~cm$^{-3}$.  
Inspecting Table~1, such a value of $N_{e,2>300keV}/N_{e,2}$ is found for a TS with a density jump of 1.12
and an Alfv\'en-Mach number of 1.09. Thus, the TS with these special parameters is able to heat
the downstream plasma from 1.55~keV up to 19.4~keV, and hence to explain the production of
relativistic electrons during the solar event on September 10, 2017.

The TS considered is a low Alfv\'en-Mach number shock. Adopting the plasma parameters in the 
upstream region of the TS as derived here (see Table~1),
the upstream flow has a velocity of $v_{TS}$ = 8342~km s$^{-1}$. At the TS, which is a fast 
magnetosonic shock, the kinetic energy of the upstream flow is transformed into an increase 
in the temperature (heating) and the magnetic field in the downstream region (see Priest (1982)).
This flow carries a total energy density of
\footnote{The thermal energy density can be neglected with respect to the other ones.}
$w_{total} = w_{kin} + w_{mag}$ = 1.027~$\times$~10$^{4}$~erg cm$^{-3}$, 
with a kinetic energy density of
$w_{kin} = N_{e,1}m_{p}v_{TS}^{2}/2$ = 5.198~$\times$~10$^{3}$~erg cm$^{-3}$ 
and a magnetic energy density of
$w_{mag} = B_{1}^{2}/8\pi$ = 5.071~$\times$~10$^{3}$~erg cm$^{-3}$.
Hence, an energy of $\approx$~641~keV is available for each electron. This result demonstrates
that the inflow in the upstream region contains enough energy 
to produce energetic electrons up to relativistic energies by means of heating 
at a low Alfv\'en-Mach number shock (such as $M_{A}$ = 1.09, for instance). 

To illustrate this, Figure 2 shows the differential electron number density
of this special TS in the downstream (full line) and upstream regions (dashed line).
It reveals that the electrons are strongly energized up to relativistic energies
at the transition through the TS. There is a pronounced part of relativistic energies
in the downstream region in the event under study. 58~\% of all electrons reach energies 
beyond 20~keV (see the eighth column in Table~1). This result agrees with the EOVSA observations 
(Fleishmann et al. 2022 ) and confirms that the energetic electrons carry a substantial part 
of the energy released during flares 
(Lin \& Hudson 1971, 1976; Emslie et al. 2004, Krucker et al. 2010, Krucker \& Battaglia 2014). 
\begin{figure}[t]
   \includegraphics[width=0.48\textwidth]{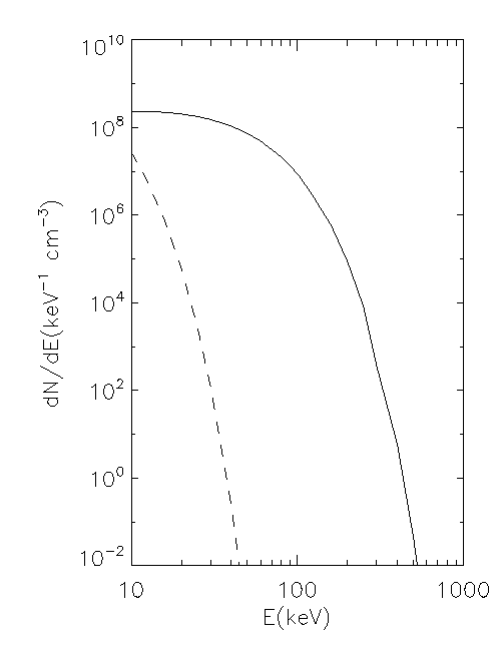}
   \caption{Energy dependence of the differential electron number density, $dN_{e}/dE$, 
           for plasma with a thermal energy of 1.55~keV 
           ($\epsilon_{th,1}$ = 0.0030 and $N_{e,1}$ = 8.929$\times$10$^{9}$~cm$^{-3}$; dashed line) 
           and 19.4~keV ($\epsilon_{th,2}$ = 0.0379 and $N_{e,2}$ = 10$^{10}$; full line) according to Eq.~(8).}
\label{Fig.2}
 \end{figure}
Chen et al. (2021) derived electron energy spectra for the event on September 10, 2017
by means of a combined analysis of hard X-ray and microwave data measured by RHESSI
(Lin et al. 2002) and EOVSA (Gary et al. 2018), respectively. Figure~3c in Chen et al. (2021)
shows the differential electron number density (thick black curve) for the impulsive phase 
of this event. It consists of a thermal (i.e., Maxwellian) core and a broken power law
in the energetic tail. Figure 3c in Chen et al. (2021) and Figure~2 of this paper 
are presented on the same scale. 
The comparison of these figures reveals that the differential electron number density
resulting from the heating at the TS (see full line in Fig.~2) agrees roughly with the
differential electron density obtained from the observations (thick black curve
of Fig.~3c in Chen at al. (2021)). We note that Chen et al. (2021) derived a broken power law 
for the differential electron number density, whereas the heating at the TS results 
in a Maxwellian EDF. (With respect to thís subject, we refer to the  discussion 
in Section~2.)


Previously, Mann et al. (2006, 2009) and Warmuth et al. (2009) proposed the generation 
of energetic electrons at the TS by SDA.
It is well known that SDA immediately produces a 
beam-like EDF (Mann et al. 2018) in the upstream region of the shock.
In contrast to these studies, here we demonstrate using the example of the
September 10, 2017 flare
that the heating of the plasma across the TS provides enough energetic
electrons up to the relativistic energies needed for the microwave and hard X-ray 
emission. The resulting EDF is not beam-like as in the case of SDA
but has a broad pronounced part at relativistic energies (see Fig.~2).



%
%

%
\section{Summary and conclusions}
There are signatures of a TS in the magnetic reconnection outflow region in solar
radio (Aurass et al. 2002, Aurass \& Mann 2004, Chen et al. 2015)
and EUV (Polito et al. 2018; Cai et al. 2021) observations as well as
numerical simulations (Forbes 1986, 1988; Forbes \& Malherbe 1986, Workman et al. 2011,
Takasao \& Shibata 2016, Takahashi et al. 2017, Shen et al. 2018, Kong et al. 2019). 
Tsuneta and Naito (1998) proposed the TS as the source of energetic electrons responsible 
for the hard X-ray loop-top sources observed in solar flares.

The X8.2 solar flare on September 10, 2017 (Chen et al. 2020) gives us a unique opportunity 
to study the generation of energetic electrons with observations from the
EOVSA instrument (Gary et al. 2018). 
This event clearly supports the standard (CSHKP) model of eruptive flares.
Signatures of a current sheet were identified by Warren et al. (2018) and
Chen et al. (2020). Relativistic electrons were produced in the vicinity of 
this current sheet. A value of 1.6 $\times$ 10$^{4}$ electrons per cm$^{3}$ with energies 
$>$~300~keV was derived from the EOVSA microwave images and spectra in the region 
between the lower end 
of the current sheet and the underlying loops; that is, the microwave emitting energetic
electrons are strongly concentrated in the loop-top region. This finding can be explained by the 
presence of a TS in the region between the lower end of the current sheet and the underlying 
flare loops. The excellent EOVSA data of this event (Chen et al. 2020) allows one to determine
the plasma parameters in the upstream and downstream regions of the TS. 
Adopting these parameters, we investigated
the jump in the temperature across the shock by means of the Rankine-Hugoniot 
relationships (see Table~1). We show that a low Alfv\'en-Mach number TS with 
$M_{A} \approx$~1.09 is able to produce enough of the relativistic electrons needed to explain 
the microwave radiation measured by the EOVSA instrument in the September 10, 2017 flare.
A substantial fraction -- namely, 58\% -- of all electrons gain an energy beyond 20~keV.

In conclusion, the generation of relativistic electrons by heating the plasma at the TS 
is a possible mechanism 
of producing energetic electrons up to the relativistic energies 
needed for microwave and hard X-ray emission 
during large solar flares. The reason is that a large ratio between the Alfv\'en velocity
and the sound speed upstream of the TS leads to a strong energizing of the downstream plasma.
\acknowledgements
GM and FS express their thanks to the financial support by
the Deutsches Zentrum f\"ur Luft- und Raumfahrt (DLR)
under the grant 50 OT 2304.
AMV aknowledges the Austrian Science Fund (FWF) under the grant 10.55776/14555.
\end{document}